\documentclass[conference]{IEEEtran}

\usepackage{cite}
\usepackage{amsmath,amssymb,amsfonts}
\usepackage{algorithmic}
\usepackage{graphicx}
\usepackage{textcomp}
\usepackage{xcolor}
\usepackage[hidelinks]{hyperref}
\def\BibTeX{{\rm B\kern-.05em{\sc i\kern-.025em b}\kern-.08em
    T\kern-.1667em\lower.7ex\hbox{E}\kern-.125emX}}

\usepackage{amsmath}
\usepackage{graphicx}

\makeatletter

\def\ps@IEEEtitlepagestyle{%
  \def\@oddfoot{\mycopyrightnotice}%
  \def\@evenfoot{}%
}
\def\mycopyrightnotice{%
  {\footnotesize 978-1-6654-9106–8/22/\$31.00~\copyright~2022 IEEE\hfill}
  \gdef\mycopyrightnotice{}
}

\@ifundefined{showcaptionsetup}{}{
 \PassOptionsToPackage{caption=false}{subfig}}
\usepackage{subfig}
\makeatother

\usepackage{eso-pic}

\title{Improving Segmentation of Breast Ultrasound Images: Semi Automatic Two Pointers Histogram Splitting Technique\\
}

\begin{document}

\author{\IEEEauthorblockN{Rasheed Abid}
\IEEEauthorblockA{\textit{Biomedical engineering} \\
\textit{Illinois Institute of Technology}\\
Chicago, IL, USA \\
r.abid94.bogra@gmail.com}
\and
\IEEEauthorblockN{S. Kaisar Alam}
\IEEEauthorblockA{\textit{Rutgers University}\\
Newark, NJ,USA \\
kaisar.alam@ieee.org}
}

\maketitle

\begin{abstract}
Automatically segmenting lesion area in breast ultrasound (BUS) images is a challenging one due to its noise, speckle and artifacts. Edge-map of BUS images also does not help because in most cases the edge-map gives no information whatsoever. Almost all segmentation technique takes the edge-map of the image as its first step, though there are a few algorithms that try to avoid edge-maps as well. Improving the edge-map of breast ultrasound images theoretically improves the chances of automatic segmentation to be more precise. In this paper, we propose a semi-automatic technique of histogram splitting using two pointers. Here the user only has to select two initially guessed points denoting a circle on the region of interest (ROI). The method will automatically study the internal histogram and split it using two pointers. The output BUS image has improved edge-map and ultimately the segmentation on it is better compared to regular segmentation using same algorithm and same initialization. Also, we further processed the edge-map to have less edge-pixels to area ratio, improving the homogeneity and the chances of easy segmentation in the future. 

\end{abstract}


\section{INTRODUCTION}
Breast cancer is the second most deadly cancer for women. In US only, it was estimated that in 2018-2019, every 1 of 8 US women will develop invasive breast cancer, over 268,600 new cases ~\cite{Zhang2022-rg}. The statistics of the previous year 2017-2018, the number of death was 40,610 ~\cite{Pons2013-bv}. This makes breast cancer study for ultrasound imaging a pressing need for the researches. Irteza et. al. ~\cite{7835383},~\cite{Kabir2022-lq}  described getting strain images from breast RF data, which can be noisy due to motion of the patient. Getting segmentation maps along with the strain images will help the radiologist to diagnose the tumor. 

Segmentation is a challenge when applied on ultrasound images. ~\cite{Noble2006-eu} surveys and describes about segmentation algorithms used on Breast Ultrasound images. ~\cite{Mukaddim2016}  mentioned methods to find out seed point in breast ultrasound images. ~\cite{Cheng2010-iy},~\cite{Shan2008-gr} mentions that due to poor quality images and noise, speckle and artifacts, it is relatively harder to find.  Many characteristics dependent segmentation algorithms like the automatic ~\cite{Xian2015-md} or semi-automatic ~\cite{Zhou2014-fb} or neural networks such as described in [ref6] have tried to segment the Region of interest (ROI). Still very high accuracy is not achieved because the properties of breast tissues ~\cite{Ramiao2016-kl} plays a big role in making shadows and low-resolution imaging make it harder for the algorithm to understand the image. As a first step towards segmentation, most algorithms take the edge-map of the image to get an approximate idea about the gradient change. For ultrasound breast image, the edge image provides almost no clear information at all. This is due to the fact that our modern edge determining algorithms find edges that are very sensitive in nature. Even with slightest of sharp gradient change, these algorithms such as Canny edge detector ~\cite{Nikolic2016-ol}, improved Sobel edge detector ~\cite{Gao2010-ol} can find edges. In an ultrasound image this sensitive property of the edge finding algorithm works as a double edged sword. It figures out the ROI but due to unwanted noise, which is many times more in ultrasound image than in regular image, the edge finding algorithms also considers the noise and artifacts to be edges. Regular edge finding algorithms fail to distinguish between ROI and other artifacts because of low signal to noise ratio of ultrasound images. Improving the edge image around the lesion area i.e. clearing out unnecessary noise edges, should help us do better segmentation with traditional algorithms, at least theoretically. In this paper, we propose a semi-automatic technique, that improves the edge-map around the lesion area of a BUS image, improving the segmentation precision with regular algorithms.

Improving the edge images require manually initializing two points on the original image, which will cover a major region of the ROI and from there the method will initialize a simple shape automatically. This shape is used to determine estimated pixel information of the ROI and then two separating pointers are used on its histogram to filter the main image. This filtered image, which we call histogram split and stretched image (HSSI) has a better edge-map with respect to the original edge-map. General segmentation algorithms such as active contour ~\cite{Kass1988-nc}, level set ~\cite{Li2010-eq} and MetaMorph ~\cite{Huang2008-hh} segmentation has yielded better results on these intermediate images. To measure the improvement, we needed the ground truth of the original BUS images, which were manually segmented by professional radiologists and in most cases, codes developed by the authors themselves. Finding the benchmark approach to quantify the results, Dice Similarity Co-efficient of the segmented intermediate image versus its ground truth was obtained and then compared with its original counterpart to find the results, methods that was described in ~\cite{Zhang2022-rg},~\cite{Pons2013-bv},~\cite{Pons2015} and also partially in ~\cite{Zhou2014-fb}.

Many machine learning algorithms ~\cite{Hosain2022-ty},~\cite{Hosain2022-rr} are being used to diagnose disease from medical images, not only with ultrasound, but also with photo acoustic imaging ~\cite{Zheng2021-uy} and MRI ~\cite{Mann2019-nz}. The 2-pointer method developed in the paper might help the diagnosis process by delineating anomaly in the image from background.

\section{Methodology}
Being a semi-automatic process, the method requires the first step of manual selection by the user. The user has to select two points to initialize a simple shape within the ROI, from which the histogram is plotted and after that image processing automatic steps take place. 

\subsection{Manual Initialization}
Motivation of this step is to have estimating information about the lesion. Usually breast cancer lesions have deformed shapes and sizes ~\cite{Ramio2016}. It is very difficult to categorize them into any common classes. To extract any feature from the lesion, we need to manually put a simple and uniform shape, such as a circle inside the lesion. An user puts two points, the center of the circle and a point on the circumference respectively, which has the radius (R) computed automatically from the coordinates, to generate this initial circle. Reason for having a circle instead of a square or rectangle or any other shape is that the trajectory of the second point is equidistant from the first, making it easier to estimate during selection. The circle should be within the lesion, covering as much as possible. Slightly overshoot-ed circles don’t show much varied result than the fully engulfed (within the lesion) one. This is due to the fact that in most of the cases, a small change in shape of the circle does not make big changes in its histogram outcome. Also, for most segmentation algorithms, we have to initialize many points. Here we manually initialize only the two points, the circumference, which is generated automatically, feeds the other points into the algorithms, which will follow later. 

\subsection{Processing the image}
\subsubsection{Processing the imageSetting up the parameters}
Different parameters were used in the processing part. To obtain the optimal result in general, brute-force implementation followed by pixel irregularities test (mentioned elaborately in later section) was done on all the images. Gray-scale being from 0 to 255, and with two-pointers to put into place, we obtained 255x254 images with pointers set at all possible place of the scale. Following processing steps will mention and use the parameters that were found to be optimal for the whole data set.

\subsubsection{Setting Two-Pointers}
Plot the histogram of the pixels within the initialized circle. In an ideal scenario, a lesion area should have only one valued pixels. In practice, artifacts, noise, speckles add up different pixels and saturation level is not constant as well. But in general, there will be dominance of the original pixels than the noisy ones. Looking at the histogram, we will point out the Most Contributing Pixels (MCP) for the lesion and the Least Contributing Pixels (LCP) that seem to appear due to practical reasons. LCP is one of the major causes of edge-pixels that appear in the edge image of the BUS image. As most segmentation algorithms look into the edge-maps of an image, these unwanted edges for the breast image plays a major role for imperfect segmentation. To isolate MCP from LCP, we apply two-pointers on the obtained histogram. The algorithm follows:- 

1. If the histogram is of the form of exponentially decaying or similar, that is the lesion area is almost saturated with black pixels. 
\begin{itemize}
    \item Put the left pointer (LP) at pixel xl = 0, that is, 
    LP←xl=0
    \item Put the Right Pointer (RP) at, xl+1, that is, RP←xr+1
    \item Keep sliding the RP to the right and stop at index xr such that, Vxr $<$ RightThreshold×V(xp)
\end{itemize}

where, xp is the pixel index with the maximum peak in the histogram, and Vi is the number of pixels at the index i and RightThreshold is the value that denotes the border between MCP and LCP on the right of maximum peak. Evaluating from section 2.2.1 that on an average LCP cannot be more that 10\% of the peak values, we came to obtain that RightThreshold = 0.10 works the most efficiently, though, changing the values differ in results as well. We will mention the same notations from here on. 
2. If the histogram has a form of Bell Curve, this means the lesion area is homogeneous with pixels of same statistics. In that case the algorithm follows. 

\begin{itemize}
    \item Firstly, find the maximum peak V(xp), where xp is the index of that peak location in the histogram. 
    \item Put the RP at (xp +1) and LP at (xp -1). That is,LP←xp-1; RP←xp+1  
    \item Keep sliding the RP to the right and stop at the index xr such that, 
    Vxr$<$RightThreshold×V(xp)
    Where, RightThreshold=0.10 
    \item   Keep sliding the LP to the left and stop at the index xl such that, 
    Vxl$<$LeftThreshold×V(xp)
\end{itemize}

where LeftThreshold denotes the value of the border between MCP and LCP on the left side of the peak. From section 1(B), we obtained LeftThreshold = 0.25 works the most efficiently for most of the cases. Reason for RightThreshold to be relatively less than LeftThreshold is that the fact that lesions with homogeneous pixels tend do have pixel values near zero and the noise usually have whiter values. If there is no such point that satisfies the condition, we stop at xl = 0 as we have reached the minimum border. If the histogram comes up with any other format then that is mentioned above, we follow the algorithm of Bell Curve. This ideally works because usually the peak and its nearby points represent a significant amount of points in the ROI.
\subsection{Histogram Splitting}
\begin{figure}[htp]
    \centering
    \includegraphics[width=8cm]{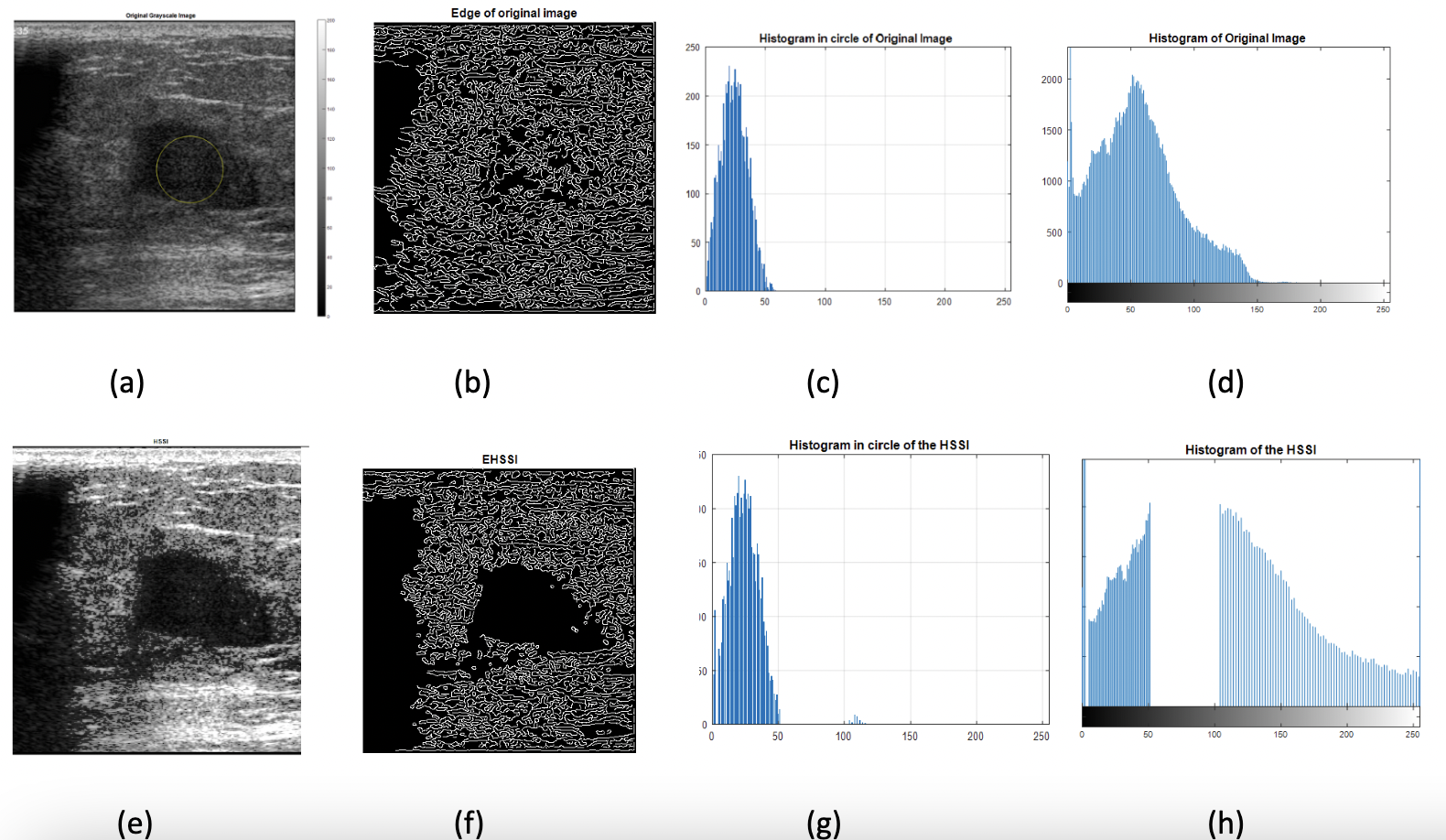}
    \caption{(a) Original Image (b) Edge map (c) Histogram within the initialized circle (d) Histogram of the whole image (e) HSSI (f) EHSSI (g) Histogram of HSSI within the initialized circle (h) histogram of HSSI}
    \label{fig:Fig1.1}
\end{figure}

It is very likely that the pixel characteristics within the lesions are quite different than the pixels outside that area. Using this fact we use RP and LP to process the whole image to get a better filtered image. We call this filtered image Histogram Split and Stretched Image (HSSI). To get this HSSI we take all the pixel values that is right to RP on the histogram indices, xi>xr, and multiply it with a RightSplittingFactor (RSF), which is greater than 1. The reverse is implemented on the pixel values of xi<xl, that is left to LP, multiplying with a LeftSplittingFactor (LSF), which is less than 1. The splitting of these pixels have a meaningful effect on the real image, providing a new insight of the HSSI. Looking into the histogram in figure 1(g) and 1(h), we notice that the histogram is split and later on stretched depending on the RSF and LSF, hence the name of the image. It makes a blank region between the MCP and LCP, both in the histogram of within the circle and the whole image. Pixels at LCP change and this makes a significant relative difference between the pixels values within the ROI and outside the ROI (Figure 1(e) ). The edge image of this HSSI at figure 1(f) has better edge image. In most cases it is clear of many edges within the lesion area. The difference between the edge image of the original BUS image in figure 1(b) and the edge image of the HSSI (EHSSI) in Figure 1(f) is clearly visible.


\subsection{Segmentation}
To determine how classical segmentation algorithms perform on with respect to original image, we used active contour snake model ~\cite{Kass1988-nc}, Metamorph segmentation ~\cite{Huang2008-hh}, Regular Level Set and Distance Regularized level set evolution segmentation model ~\cite{Li2010-eq}. All of these models take in some initialized points, to estimate ROI. In this case, we use the unmodified version of these algorithms on both the original image and HSSI with the same initialized points. After the segmentation we obtain the border and derive the binary image from it, white being the object, black the background.

\subsection{Morphological Operations}
This is a work in progress, which shows statistical improvement in the process, which may be used in the future to develop segmentation algorithms just from the edge image. 
Looking into the edge image of the HSSI (EHSSI) we can see the improvement when compared with the original edge image.  Split histogram made the lesion area more saturated, leaving less edge pixels, and more noise outside the ROI, adding to previous edge pixels. We refer “edge pixels” to those pixels which appear in as pixels in edge image. These change in the edge pixels help the segmentation algorithms perform better, as most of the algorithms, take edge map of an image as an initial parameter to find the ROI. 
One probable target of work in the future can be to only use this edge map to segment out the ROI of the original image. For this, we perform two operations on the EHSSI –

1. Edge wash-up: It is obvious that the center of the circle must be in the lesion. To remove unwanted edge pixels that appear very close to the center we wash up all the edge pixels in a small radius (Edge wash-up radius, EWR) from the center of the circle. The size of EWR, depends on the radius (R) of the initialized circle, proportionally. The obtained image is called “Washed-up” image. We perform our pixel irregularities test on the result section to find out more about image. 

2. Imclose operation: To fill up the holes, we perform flood-fill operation in the background (imfill in Matlab image processing toolbox) followed by dilution and erosion and lastly flood-fill again. 
This process tends to make the outside of the ROI look like a solid background, we call this image Enhanced EHSSI or EEHSSI. We will follow up the results of the precision of EHSSI in the result section. 

\begin{figure}[htp]
    \centering
    \includegraphics[width=8cm]{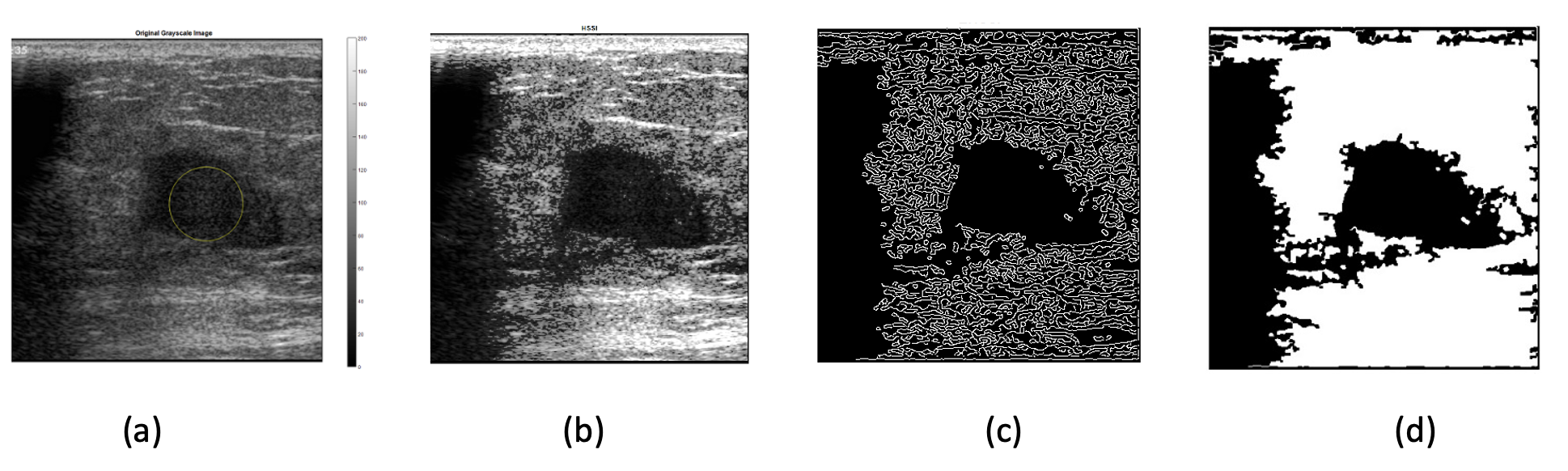}
    \caption{Figure 2. (a) Original Image (b) HSSI (c) EHSSI (d) EEHSSI}
    \label{fig:Fig2.1}
\end{figure}

\section{ Dataset}
The study was compared on a dataset of 221 generated B-mode images, acquired from 94 patients from ATL’s PMA study, which was conducted in 1994 ~\cite{Gao2010-ol}. This dataset was taken from the study in ~\cite{Mukaddim2016}. To implement the study, we used a personal computer with windows operating system. We used Matlab2016a with a hardware of Intel® Core™i5-4570 CPU with 3.20GHz clock cycle, 8GB of RAM and a GPU of Nvdia GTX 1050ti.  


\section{Results}
Two different tests were conducted to find out the quantitative improvement. Dice similarity co-efficient test – to find out improvement in segmentation algorithms; and, pixel irregularity test – to find out the improvement of edge pixels in EHSSI and EEHSSI.

\subsection{Dice Similarity Co-efficient}
All the ground truth of the BUS images were manually delineated by expert clinicians. Comparing the segmented regions by the algorithms with the ground truth image we can see improvement in the results. As a comparison parameter, we used Sorensen Dice Similarity Co-efficient (DSC). Many previous works such as ,~\cite{Zhou2014-fb} ~\cite{Zhang2022-rg}, ~\cite{Pons2013} etc. have used the following DSC formula to determine how precise the segmentation is with respect to the manually delineated one. If the segmented binary image is X (fig. 3(c)) and the ground truth is Y fig. (3(d)) , then, DSC parameter

\begin{equation}
\mathrm{DSC}=\frac{2^*|X \cap Y|}{|X|+|Y|}
\end{equation}

Segmentation algorithms of the study were collected from various source. To compare the results with active contour, we used Matlab inbuilt function activecontour [Matlab copyright 2012-2017] for the active contour segmentation. For MetaMorph segmentation, we used the codes provided by the authors themselves. Two level set methods, Distance Regularized Level Set Evolution (DRLSE) and Spacial Fuzzy Cluster and Level Set Segmentation (SFCLS) were collected from MathWorks File Exchange, uploaded by the author himself for DRLSE at ~\cite{Li2010-eq} and SFCLS by “ABing” from National University of Singapore in ~\cite{Li2011-fn}.  

\begin{figure}[htp]
    \centering
    \includegraphics[width=8cm]{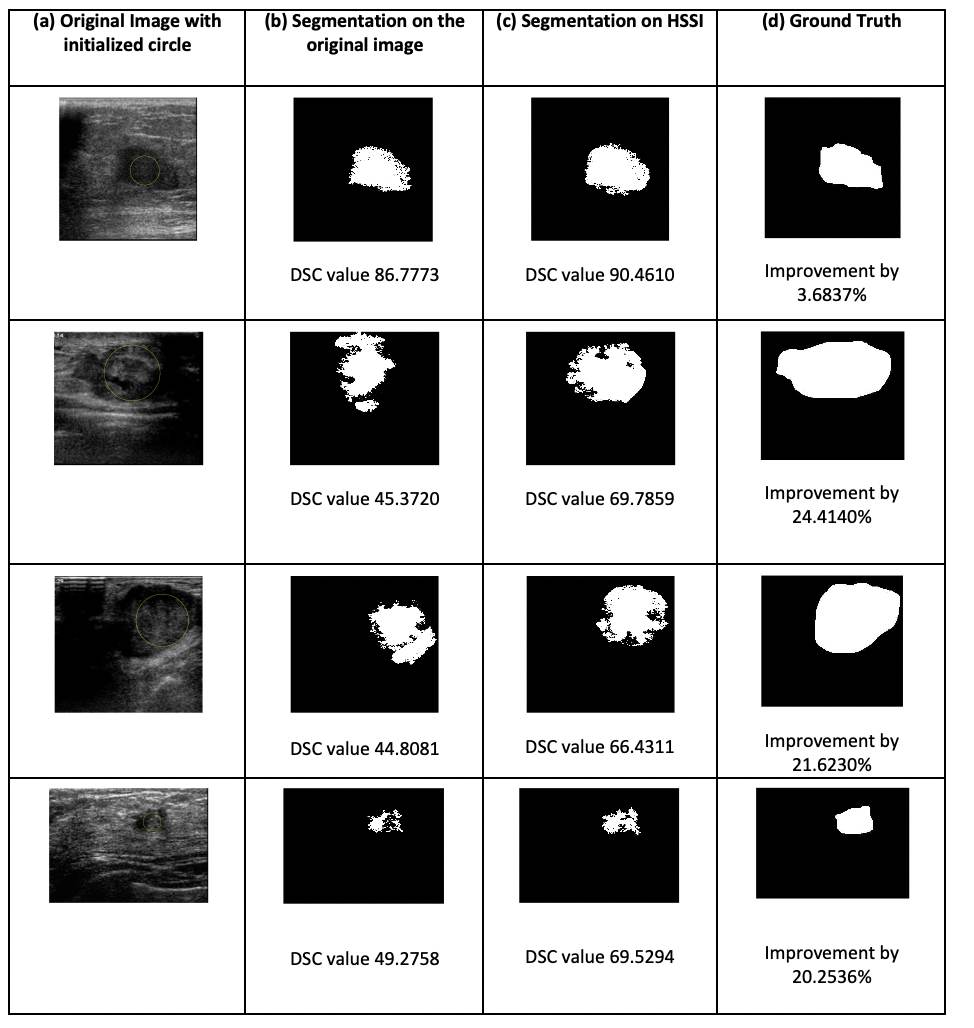}
    \caption{A visual representation of the binary images and their DSC values along with improvement (a) Original image with the initialized circle (b) Segmentation result on the unmodified image (c) Segmentation result on HSSI (d) ground truth}
    \label{fig:Fig3}
\end{figure}

Comparing with different segmentation algorithms, we get the following results:

\begin{table}[]
\centering
\caption{A comparison of Dice Similarity Co-efficient between segmentation of original image and HSSI using (a) active contour segmentation (b) MetaMorph segmentation (c) Distance Regularized Level Set Evolution (d) Spatial Fuzzy Clustering and Level Set Segmentation and their number of success.}
\label{tab:comparison}
\begin{tabular}{|l|l|l|l|}
\hline
DSC (\%) & Original Image                                                                                      & HSSI                                                             & Success (221 images)                                                   \\ \hline
Active Contour    &                                        62.65       & 65.888  & 140                                                        \\ \hline
MetaMorph  &  36.12       & 40.77  & 191

 \\ \hline
DRLSE      &  58.32 & 60.43 & 156
\\ \hline
SFCLC     &  59.21 &62.68 &147
\\ \hline
\end{tabular}
\end{table}

A thing to note is that the results of segmentation of the algorithms may seem to lag the original claimed quality. The reason is the initialization. Most algorithms require more precise initialization, where as our method only requires a not-very-precise simple circle to initialize.

\subsection{Pixel Irregularities Test }
Pixel irregularities test is a study performed on the lesion area of the edge maps obtained from the original image and the HSSI (EHSSI). In the lesion area (fig. 4(c)), ideally there should be no edge pixels, because in ideal case, that region should be homogeneous. But practical ultrasound diagnosis ends up with low-resolution, non-homogeneous regions, yielding edge pixels in the lesion area. HSSI provide us with more homogeneous region in the ROI. To compare the two edge images fig. 4(c) and (d), we perform test to find the percentage of wrong pixels appearing in the lesion area of the edge map. This percentage is the ratio of wrong pixels appearing in the ROI and the area (by pixels) of the ROI by hundred, denoted by Pixel Irregularities Ratio (PIR). Area of the ROI was obtained from the ground truth.

\begin{equation}
P I R=\frac{\text { Edge Pixels in the ROI }}{\text { Area size of the ROI }} \times 100
\end{equation}

It was also performed on the washed-up image (fig. 4(e)) with better result outcomes. As we know, the less the wrong pixels in the ROI, the lower the PIR, the better the result, showed in Table 2. We also calculated the number of images where our method yields better edge map inside the ROI.


\begin{table}[]
\centering
\caption{Comparison of PIR between Original image, HSSI and Washed-up image and the number of success.}
\label{tab:comparison}
\begin{tabular}{|l|l|l}
\hline
 & PIR                                                            & Success (Out of 224 images)                                                   \\ \hline
Original image   &                                        16.9225      & N/A                                                      \\ \hline
EHSSI  &  10.6357

      & 181

 \\ \hline
Washed-up Image      &  9.5676 & 195

\\ \hline
\end{tabular}
\end{table}

\subsection{Weak Cases}

\begin{figure}[htp]
    \centering
    \includegraphics[width=8cm]{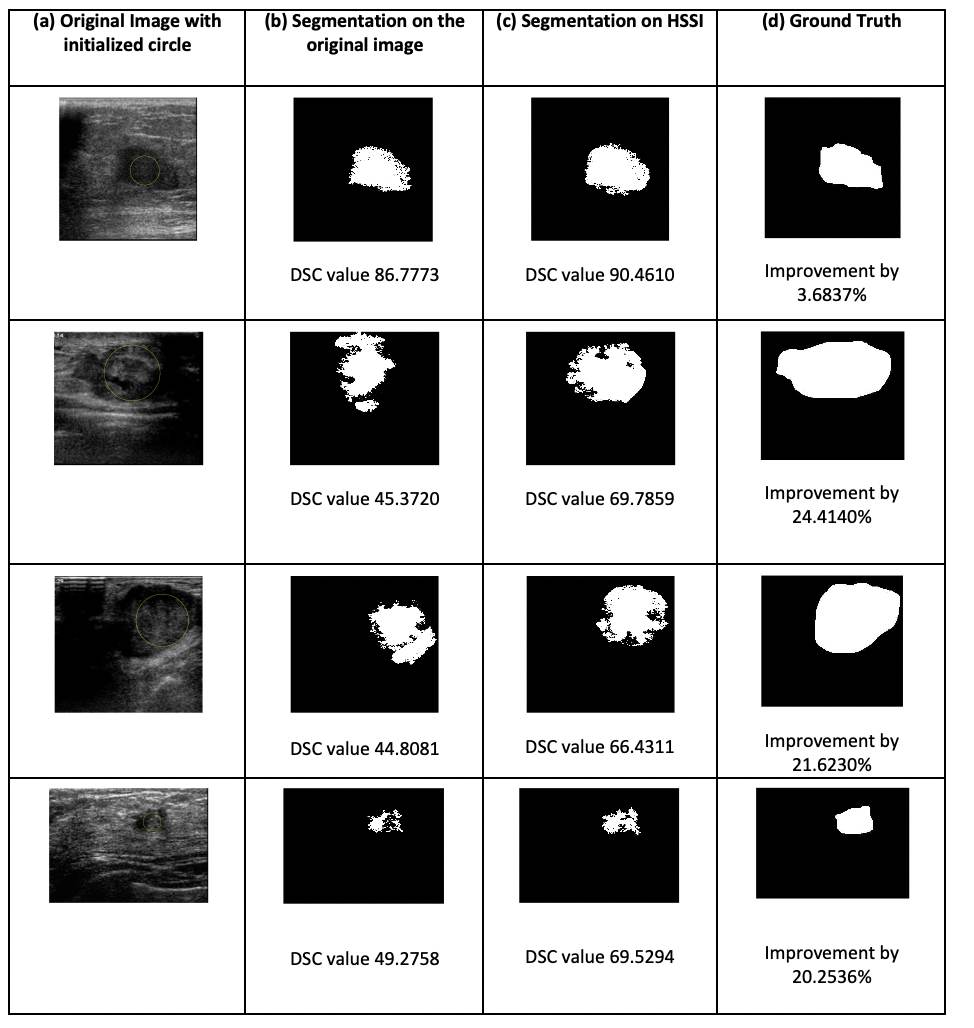}
    \caption{Weak cases (a) Original image with circle initialization (b) Original Edge Image (c) EHSSI}
    \label{fig:Fig4}
\end{figure}
At worst cases, our method can keep the system as it is. If the lesion area is very random in nature, then the two pointers are not very helpful to distinguish the MCP and LCP. So, the algorithm is unable to generate better HSSI image than the original one. Figure 4 shows the comparison of images where our method fails to improve the results. 

\begin{figure}[htp]
    \centering
    \includegraphics[width=8cm]{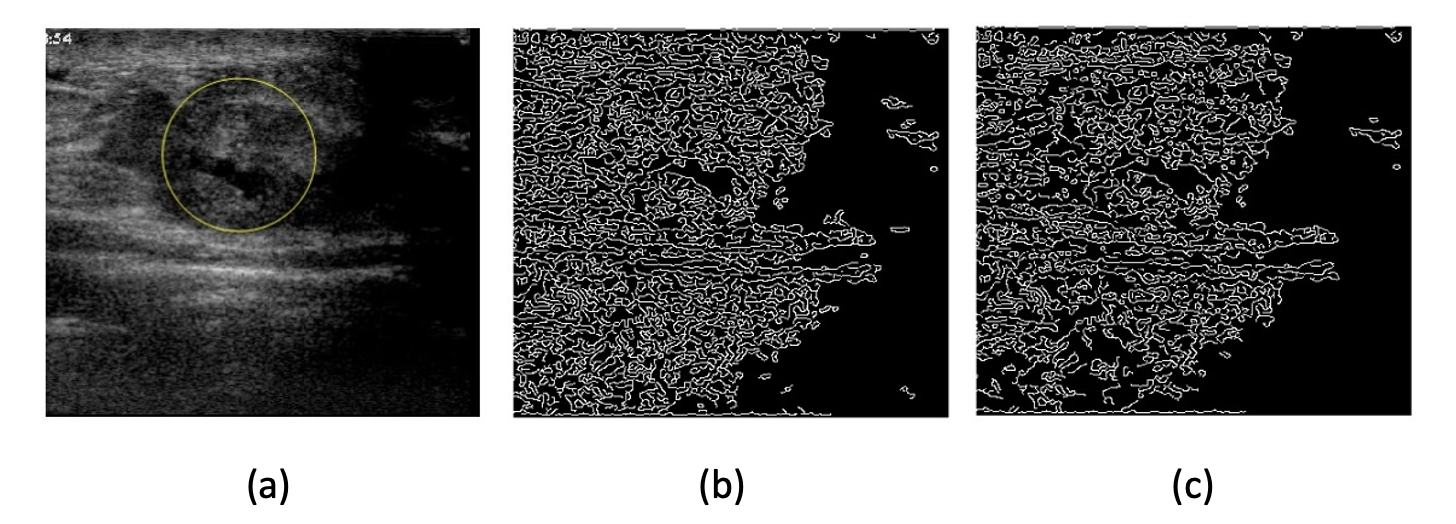}
    \caption{Weak cases (a) Original image with circle initialization (b) Original Edge Image (c) EHSSI}
    \label{fig:Fig5}
\end{figure}

\section{Conclusion}
Enhanced edge of histogram split and separated images (EEHSSI) can be used in the future to segment the lesion area only by using our method and a few morphological operations, no energy functions were required. Though we are unable to formulate something that will segment the lesion are, fully or partially, for all the cases, for a few cases we have been able to make good binary images, where the lesion area looks quite like the ground truth ROI. The improvement can be seen in figure 5.

\bibliographystyle{IEEEtran}
\bibliography{main}

\begin{thebibliography}{10}
\providecommand{\url}[1]{#1}
\csname url@samestyle\endcsname
\providecommand{\newblock}{\relax}
\providecommand{\bibinfo}[2]{#2}
\providecommand{\BIBentrySTDinterwordspacing}{\spaceskip=0pt\relax}
\providecommand{\BIBentryALTinterwordstretchfactor}{4}
\providecommand{\BIBentryALTinterwordspacing}{\spaceskip=\fontdimen2\font plus
\BIBentryALTinterwordstretchfactor\fontdimen3\font minus
  \fontdimen4\font\relax}
\providecommand{\BIBforeignlanguage}[2]{{%
\expandafter\ifx\csname l@#1\endcsname\relax
\typeout{** WARNING: IEEEtran.bst: No hyphenation pattern has been}%
\typeout{** loaded for the language `#1'. Using the pattern for}%
\typeout{** the default language instead.}%
\else
\language=\csname l@#1\endcsname
\fi
#2}}
\providecommand{\BIBdecl}{\relax}
\BIBdecl

\bibitem{Zhang2022-rg}
Y.~Zhang, M.~Xian, H.-D. Cheng, B.~Shareef, J.~Ding, F.~Xu, K.~Huang, B.~Zhang,
  C.~Ning, and Y.~Wang, ``\BIBforeignlanguage{en}{{BUSIS}: A benchmark for
  breast ultrasound image segmentation},''
  \emph{\BIBforeignlanguage{en}{Healthcare (Basel)}}, vol.~10, no.~4, p. 729,
  Apr. 2022.

\bibitem{Pons2013-bv}
G.~Pons, J.~Mart{\'\i}, R.~Mart{\'\i}, S.~Ganau, J.~C. Vilanova, and J.~A.
  Noble, ``\BIBforeignlanguage{en}{Evaluating lesion segmentation on breast
  sonography as related to lesion type},'' \emph{\BIBforeignlanguage{en}{J.
  Ultrasound Med.}}, vol.~32, no.~9, pp. 1659--1670, Sep. 2013.

\bibitem{7835383}
I.~E. Kabir, R.~Abid, A.~S. Ashik, K.~K. Islam, and S.~K. Alam, ``Improved
  strain estimation using a novel 1.5d approach: Preliminary results,'' in
  \emph{2016 International Conference on Medical Engineering, Health
  Informatics and Technology (MediTec)}, 2016, pp. 1--5.

\bibitem{Kabir2022-lq}
I.~E. Kabir, ``Imrpoving strain estimation in breast ultrasound images using
  novel 1.5d approach (simulation and in-vivo results,'' 2022.

\bibitem{Noble2006-eu}
J.~A. Noble and D.~Boukerroui, ``\BIBforeignlanguage{en}{Ultrasound image
  segmentation: a survey},'' \emph{\BIBforeignlanguage{en}{IEEE Trans. Med.
  Imaging}}, vol.~25, no.~8, pp. 987--1010, Aug. 2006.

\bibitem{Mukaddim2016}
\BIBentryALTinterwordspacing
R.~A. Mukaddim, J.~Shan, I.~E. Kabir, A.~S. Ashik, R.~Abid, Z.~Yan, D.~N.
  Metaxas, B.~S. Garra, K.~K. Islam, and S.~K. Alam, ``A novel and robust
  automatic seed point selection method for breast ultrasound images,'' in
  \emph{2016 International Conference on Medical Engineering, Health
  Informatics and Technology ({MediTec})}.\hskip 1em plus 0.5em minus
  0.4em\relax {IEEE}, Dec. 2016. [Online]. Available:
  \url{https://doi.org/10.1109/meditec.2016.7835370}
\BIBentrySTDinterwordspacing

\bibitem{Cheng2010-iy}
H.~D. Cheng, J.~Shan, W.~Ju, Y.~Guo, and L.~Zhang,
  ``\BIBforeignlanguage{en}{Automated breast cancer detection and
  classification using ultrasound images: A survey},''
  \emph{\BIBforeignlanguage{en}{Pattern Recognit.}}, vol.~43, no.~1, pp.
  299--317, Jan. 2010.

\bibitem{Shan2008-gr}
J.~Shan, H.~D. Cheng, and Y.~Wang, ``A novel automatic seed point selection
  algorithm for breast ultrasound images,'' in \emph{2008 19th International
  Conference on Pattern Recognition}.\hskip 1em plus 0.5em minus 0.4em\relax
  IEEE, Dec. 2008.

\bibitem{Xian2015-md}
M.~Xian, Y.~Zhang, and H.~D. Cheng, ``\BIBforeignlanguage{en}{Fully automatic
  segmentation of breast ultrasound images based on breast characteristics in
  space and frequency domains},'' \emph{\BIBforeignlanguage{en}{Pattern
  Recognit.}}, vol.~48, no.~2, pp. 485--497, Feb. 2015.

\bibitem{Zhou2014-fb}
Z.~Zhou, W.~Wu, S.~Wu, P.-H. Tsui, C.-C. Lin, L.~Zhang, and T.~Wang,
  ``\BIBforeignlanguage{en}{Semi-automatic breast ultrasound image segmentation
  based on mean shift and graph cuts},''
  \emph{\BIBforeignlanguage{en}{Ultrason. Imaging}}, vol.~36, no.~4, pp.
  256--276, Oct. 2014.

\bibitem{Ramiao2016-kl}
N.~G. Rami{\~a}o, P.~S. Martins, R.~Rynkevic, A.~A. Fernandes, M.~Barroso, and
  D.~C. Santos, ``\BIBforeignlanguage{en}{Biomechanical properties of breast
  tissue, a state-of-the-art review},'' \emph{\BIBforeignlanguage{en}{Biomech.
  Model. Mechanobiol.}}, vol.~15, no.~5, pp. 1307--1323, Oct. 2016.

\bibitem{Nikolic2016-ol}
M.~Nikolic, E.~Tuba, and M.~Tuba, ``Edge detection in medical ultrasound images
  using adjusted canny edge detection algorithm,'' in \emph{2016 24th
  Telecommunications Forum ({TELFOR})}.\hskip 1em plus 0.5em minus 0.4em\relax
  IEEE, Nov. 2016.

\bibitem{Gao2010-ol}
W.~Gao, X.~Zhang, L.~Yang, and H.~Liu, ``An improved sobel edge detection,'' in
  \emph{2010 3rd International Conference on Computer Science and Information
  Technology}.\hskip 1em plus 0.5em minus 0.4em\relax IEEE, Jul. 2010.

\bibitem{Kass1988-nc}
M.~Kass, A.~Witkin, and D.~Terzopoulos, ``\BIBforeignlanguage{en}{Snakes:
  Active contour models},'' \emph{\BIBforeignlanguage{en}{Int. J. Comput.
  Vis.}}, vol.~1, no.~4, pp. 321--331, Jan. 1988.

\bibitem{Li2010-eq}
C.~Li, C.~Xu, C.~Gui, and M.~D. Fox, ``\BIBforeignlanguage{en}{Distance
  regularized level set evolution and its application to image segmentation},''
  \emph{\BIBforeignlanguage{en}{IEEE Trans. Image Process.}}, vol.~19, no.~12,
  pp. 3243--3254, Dec. 2010.

\bibitem{Huang2008-hh}
X.~Huang and D.~N. Metaxas, ``\BIBforeignlanguage{en}{Metamorphs: deformable
  shape and appearance models},'' \emph{\BIBforeignlanguage{en}{IEEE Trans.
  Pattern Anal. Mach. Intell.}}, vol.~30, no.~8, pp. 1444--1459, Aug. 2008.

\bibitem{Pons2015}
\BIBentryALTinterwordspacing
G.~Pons, J.~Mart{\'{\i}}, R.~Mart{\'{\i}}, S.~Ganau, and J.~A. Noble,
  ``Breast-lesion segmentation combining b-mode and elastography ultrasound,''
  \emph{Ultrasonic Imaging}, vol.~38, no.~3, pp. 209--224, Jun. 2015. [Online].
  Available: \url{https://doi.org/10.1177/0161734615589287}
\BIBentrySTDinterwordspacing

\bibitem{Hosain2022-ty}
A.~K. M.~S. Hosain, M.~Islam, M.~H.~K. Mehedi, I.~E. Kabir, and Z.~T. Khan,
  ``Gastrointestinal disorder detection with a transformer based approach,''
  2022.

\bibitem{Hosain2022-rr}
A.~K. M.~S. Hosain, M.~H.~K. Mehedi, and I.~E. Kabir, ``{PCONet}: A
  convolutional neural network architecture to detect polycystic ovary syndrome
  ({PCOS}) from ovarian ultrasound images,'' 2022.

\bibitem{Zheng2021-uy}
E.~Zheng, H.~Zhang, S.~Goswami, I.~E. Kabir, M.~M. Doyley, and J.~Xia,
  ``\BIBforeignlanguage{en}{Second-generation dual scan mammoscope with
  photoacoustic, ultrasound, and elastographic imaging capabilities},''
  \emph{\BIBforeignlanguage{en}{Front. Oncol.}}, vol.~11, p. 779071, Nov. 2021.

\bibitem{Mann2019-nz}
R.~M. Mann, N.~Cho, and L.~Moy, ``\BIBforeignlanguage{en}{Breast {MRI}: State
  of the art},'' \emph{\BIBforeignlanguage{en}{Radiology}}, vol. 292, no.~3,
  pp. 520--536, Sep. 2019.

\bibitem{Ramio2016}
\BIBentryALTinterwordspacing
N.~G. Rami{\~{a}}o, P.~S. Martins, R.~Rynkevic, A.~A. Fernandes, M.~Barroso,
  and D.~C. Santos, ``Biomechanical properties of breast tissue, a
  state-of-the-art review,'' \emph{Biomechanics and Modeling in
  Mechanobiology}, vol.~15, no.~5, pp. 1307--1323, Feb. 2016. [Online].
  Available: \url{https://doi.org/10.1007/s10237-016-0763-8}
\BIBentrySTDinterwordspacing

\bibitem{Li2011-fn}
B.~N. Li, C.~K. Chui, S.~Chang, and S.~H. Ong,
  ``\BIBforeignlanguage{en}{Integrating spatial fuzzy clustering with level set
  methods for automated medical image segmentation},''
  \emph{\BIBforeignlanguage{en}{Comput. Biol. Med.}}, vol.~41, no.~1, pp.
  1--10, Jan. 2011.

\end{thebibliography}

\end{document}